\documentclass[
               biblatex,      
               ]{jacow}

\usepackage[english]{babel}

\ExplSyntaxOn
\cs_if_exist:NF \tl_set:Ne          { \cs_generate_variant:Nn \tl_set:Nn          { Ne }   }
\cs_if_exist:NF \tl_gset:Ne         { \cs_generate_variant:Nn \tl_gset:Nn         { Ne }   }
\cs_if_exist:NF \regex_match:NVTF   { \cs_generate_variant:Nn \regex_match:NnTF   { NVTF } }
\cs_if_exist:NF \regex_match:nVTF   { \cs_generate_variant:Nn \regex_match:nnTF   { nVTF } }
\cs_if_exist:NF \regex_match:nVF    { \cs_generate_variant:Nn \regex_match:nnF    { nVF }  }
\cs_if_exist:NF \regex_replace_all:NnN { \cs_generate_variant:Nn \regex_replace_all:nnN { NnN } }
\ExplSyntaxOff

\DeclareSIUnit\urad{\micro\radian}
\DeclareSIUnit\clight{\ensuremath{c}}   

\begin{document}

\title{Development of a Novel High Angular Resolution Tracking System at the CERN SPS North Area Test Beam}

\author{J. Braach\thanks{Corresponding author. Contact: justus.braach@cern.ch}, H. Sandberg, J. Storey, J. J\~{o}ul, L. M. Golino, L. S. Esposito, \\M. McLean, M. M. Bay, W. Devauchelle}
\affil{CERN, Geneva, Switzerland}

\maketitle

\begin{abstract}
   The BIPXL telescope is a high-precision tracking system developed at CERN for the characterization of bent silicon crystals in secondary particle beams. 
   The system enables the measurement of crystal alignment, track deflection and channeling efficiency.
   It consists of two \qty{10}{m} tracking arms equipped with up to eight hybrid pixel detector (HPD) planes, scintillator-photomultiplier tube (PMT) trigger stations, adjustable detector stages, and a high-precision multi-axis goniometer for crystal alignment. 
   Its configurable geometry accommodates different test scenarios depending on tracking performance requirements. 
   The current prototype setup consists of four Timepix3 HPDs as tracking detector planes and achieves a spatial and angular resolution down to single-digit-\unit{\um} and \qty{20}{\micro\radian}, respectively. 
   The HPDs are integrated using a radiation-hard readout architecture developed within the CERN Beam Instrumentation (BI) group. 
   Dedicated solutions for motion control, system coordination and online reconstruction enabled development and deployment of the full system in less than a year and stable multi-day test-beam campaigns since then. 
   First measurements demonstrate reliable operation and confirm the performance expected from simulations. 
   Still, further improvements in effective detection area and material budget are necessary to meet tracking requirements. 
   The presented prototype system represents a first step towards a low-material-budget high angular resolution telescope, with future developments based on next-generation Timepix hybrid pixel detectors, an improved readout and control architecture, and modular liquid cooling during reference-detector operation.
\end{abstract}

\section{Introduction}

Beam telescopes are the reference instruments of accelerator and particle-detector research and development. 
By reconstructing the trajectories of individual charged particles from the hits recorded in a set of precisely aligned reference planes, they predict the position and direction of a particle at the position of a device under test (DUT) placed among the detector planes~\cite{Tpx3Telescope, VELOtelescope}.

The BIPXL telescope is a high-precision tracking system developed in the Beam Instrumentation (BI) group~\cite{BIgroup} to be deployed at the CERN SPS North Area beamlines~\cite{spsNA}.
The development caters to the needs of the DECRYCE project (Development of Crystals for Collimation and Beam Extraction)~\cite{VelottiCrystalProd}.
Bent crystals steer charged particles through channeling and volume reflection and are used for crystal-assisted collimation and extraction, for the shadowing of electrostatic septa, and for future fixed-target experiments at the High-Luminosity LHC~\cite{CrystalCollimation, VelottiSeptum, VelottiShadowing}.
Their characterization in secondary particle beams verifies the essential specifications before installation. 
Key observables are the channeling efficiency as a function of the particle impact angle, the achieved track deflection (bending) angle, and the crystal torsion.
An angular resolution at the few-\unit{\urad} level for both the incoming and the outgoing particle track and a close-to \qty{100}{\percent} tracking efficiency set the defining requirements of the instrument.

The telescope prototype is based on Timepix3 hybrid pixel detectors (HPDs)~\cite{Poikela2014} read out and controlled with a radiation-hard, BI-native hardware/software architecture.
The full telescope setup integrates reference pixel detector, scintillator-PMT trigger and motorized linear/goniometer motion hardware through common software solutions. 
The beam telescope and test beam community-standard software frameworks Constellation, Corryvreckan and Allpix\textsuperscript{2}~\cite{Constellation, Corryvreckan, Allpix2} implement the components and required functionalities. 
They enabled full-system deployment and stable multi-campaign operation within a year.
This paper describes the design of the prototype, its main components and its preliminary tracking performance, and outlines the ongoing upgrade towards a low-material-budget telescope based on the Timepix4 Application-Specific Integrated Circuit (ASIC)~\cite{Llopart2022}.

\section{The Beam Telescope Prototype}

\subsection{Measurement Strategy}

\begin{figure}[!h]
   \centering
   \includegraphics[width=\columnwidth]{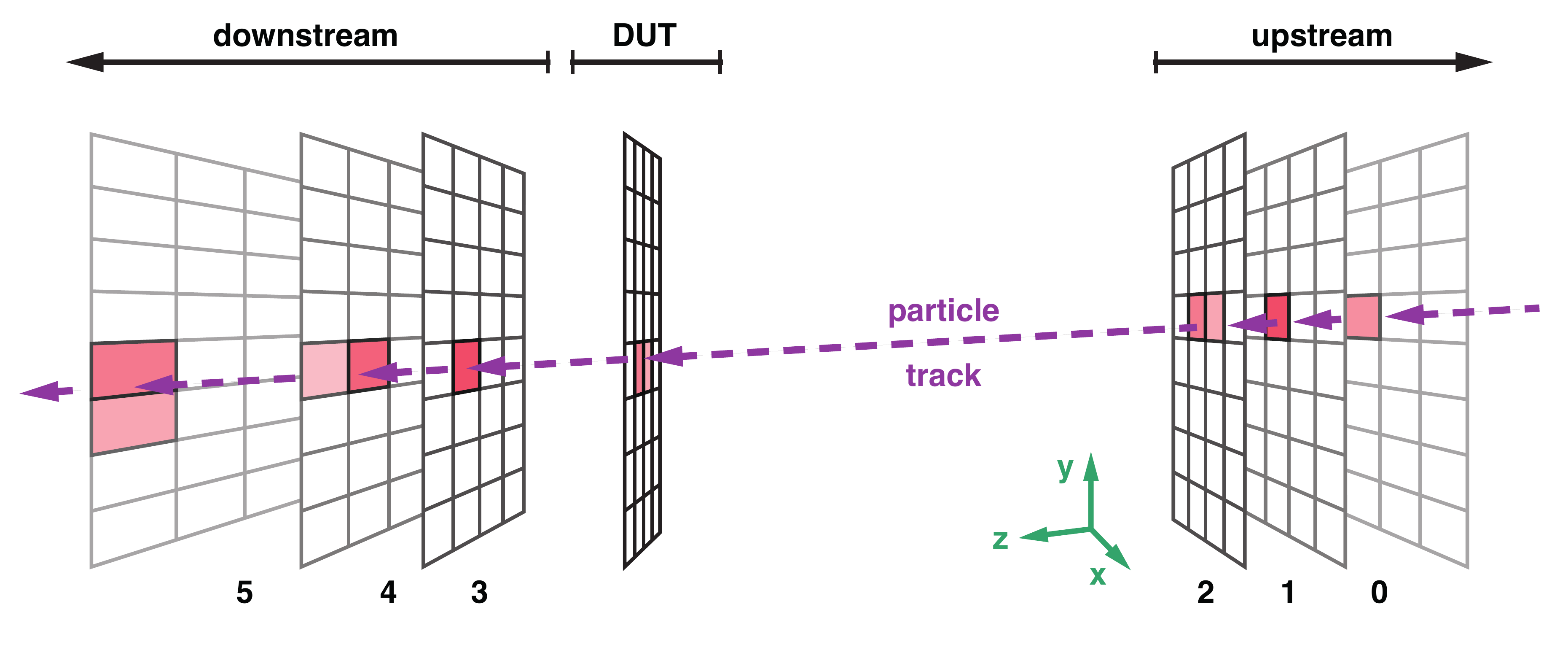}
   \caption{Core measurement principle: pixel-cluster intercepts in the upstream (0--2) and downstream (3--5) reference planes reconstruct the track of a single beam particle.}
   \label{fig:principle}
\end{figure}

The telescope reconstructs the trajectory of individual beam particles from the intercepts recorded as pixel clusters ($x$/$y$ dimension) in a set of reference detector planes arranged along the beam (z dimension), as illustrated in Fig.~\ref{fig:principle}. 
The DUT is mounted on a central goniometer between the upstream and the downstream arm of the telescope. 
Track reconstruction uses a general-broken-lines (GBL) fit.
The reference plane alignment is refined with an iterative track-based algorithm.
Both reconstruction and alignment are implemented in the Corryvreckan analysis framework~\cite{Corryvreckan}.
The deflection (kink) angle between the two track segments (tracklets) recorded by the individual telescope arms, together with the track impact position on the crystal face, yields the desired characterization observables. 
The pointing precision at the DUT combines a measurement-and-fit term $\sigma_\text{fit}$, which improves with thin, precise planes and a short lever arm, with a multiple-scattering term that grows with the lever arm $L$ and the material budget,
\begin{equation}
   \sigma_\text{DUT} \approx \sigma_\text{fit} \oplus L\,\theta_0,
   \qquad
   \theta_0 \propto \frac{1}{p}\sqrt{\frac{x}{X_0}},
   \label{eq:pointing}
\end{equation}
where $\theta_0$ is the RMS multiple-scattering angle, $p$ the beam momentum, $x/X_0$ the traversed material budget and $\oplus$ a sum in quadrature.
The deflection-angle resolution, obtained from the two tracklets, instead improves with a \emph{long} lever arm $L_\text{arm}$ until the same scattering floor dominates,
\begin{equation}
   \sigma_{\Delta\theta} \approx \sqrt{2}\,\frac{\sigma_\text{plane}}{L_\text{arm}} \oplus \theta_0,
   \label{eq:angular}
\end{equation}
with $\sigma_\text{plane}$ the single-plane spatial precision.
Spatial and angular resolution thus pull the geometry in opposite directions, requiring the plane positions and orientations to be configurable.
Spatial resolution is traded against tracking acceptance and angular resolution for a given measurement~\cite{PracticalGuide} scenario.
This opens the prototype to use cases beyond the long-baseline, high-angular-resolution configuration used here for crystal characterization.

\subsection{Telescope Layout and Mechanics}

The long-baseline telescope prototype spans a configurable baseline of about \qty{20}{m}, with lever arms of approximately \qty{10}{m} upstream and downstream of the DUT. 
Up to three reference planes can be placed at each of four longitudinal positions. 
The current prototype uses a total of four planes, located at $z\approx-11.4$, $-0.4$, $0.5$ and \qty{9.9}{m} relative to the crystal. 
Each plane is mounted on one of four foldable heavy-duty aluminum tripod towers carrying a motorized $x$/$y$ stage (Zaber LSQ series~\cite{Zaber}; travel $\geq\qtyproduct{150 x 75}{mm}$, accuracy $\lesssim\qty{30}{\um}$) and a manual yaw/tilt stage.
The former mechanically maximizes the overlap of the \qtyproduct{14 x 14}{mm} active area of the HPDs along the long-baseline setup.
The latter sets the detector inclination relative to the particle tracks, adjusting the track path length through the sensor and thus the charge sharing and spatial precision of each plane.
To save one ASIC equivalent of material in the tracking volume, upstream HPDs face ASIC-to-beam and downstream HPDs face sensor-to-beam.
The crystal is held in a C-shaped bender on the multi-axis goniometer (Physik Instrumente (PI)~\cite{PI}) at the center of the setup.
The planes are pre-aligned to about \qty{1}{\mm} with laser-metrology targets and standard surveying tools, followed by a track-based alignment down to \qty{1}{\um} precision.

The prototype telescope is installed in the H8 beam line of the SPS North Area (experimental area PPE128)~\cite{spsh8}, where it operates in a high-purity, positive-pion beam of \qty[per-mode=symbol]{180}{\giga\electronvolt\per\clight} at an intensity of about \num{1e6} particles per spill and a spot size of about \qty{3}{mm}.
The high momentum keeps the multiple-scattering contribution to the pointing resolution small (cf.\ Eq.~\eqref{eq:pointing}).
The moderate intensity keeps the per-plane occupancy low enough for unambiguous single-track reconstruction.
Vacuum pipes between the reference detectors reduce the impact of multiple scattering as particles travel along the length of the telescope arms.

\subsection{Reference Detectors}

The reference detector plane comprises a \qty{300}{\um} thick p-on-n silicon sensor bump-bonded to a Timepix3 ASIC, with a \qty{55}{\um} square pixel pitch and \qty{1.5625}{ns} time bin. 
The detectors record the position and the arrival time of each hit (4D tracking) and are read out in a data-driven, trigger-less mode, which is well suited to the high and non-uniform particle rate of the North Area spill structure.  
The assemblies and their radiation-hard readout were available within the BI group as a by-product of its beam-gas-ionization instrumentation program~\cite{BGI}.
Each plane presents a material budget of about \qty{3.5}{\percent} of a radiation length ($X_0$), dominated by the ASIC and the carrier printed circuit board (PCB).
Their specification is not ideal in every respect for high-angular-resolution tracking, but demonstrates that a capable telescope can be built on BI-native technology.

\subsection{Time Reference}

A pair of scintillator--PMT stations (Luxium BC-418 plastic scintillator~\cite{BC418} on an acrylic fishtail light guide, read out by Hamamatsu H10721P-210 PMT modules~\cite{Hamamatsu}) provides an independent coincidence trigger timestamp for each beam particle.
The timestamp is recorded by the readout back-end and is used to define a reconstruction window of about \qty{5}{\micro\second} around the trigger, within which the pixel data of a particle intercept are expected. 
Per-track time is computed as the mean of the arrival times of the pixel clusters that compose the track.
While such an external time reference is typically used to characterize a DUT's timing performance, here it instead decreases the single-track reconstruction time and can equally serve as the timing reference for the pixel-detector planes themselves.

\subsection{Readout, Services and Control}

The HPDs are read out utilizing the BI-native, radiation-hard architecture described in~\cite{ReadoutTpx3, NextGenReadout}.
The readout front-end is based on CERN's Gigabit Transceiver (GBTx) chips and radiation-hard FEASTMP DC/DC converters, and drives the bidirectional optical link through Versatile Transceivers (VTRx)~\cite{ReadoutTpx3}.
On the opposite end, an FPGA in the back-end crate receives and processes the data and facilitates detector control.
An industrial computer collects the event packets and writes them to disk in a BI-native HDF5 format.
A programmable logic controller (PLC) handles slow control: separate power domains for the front-end board, pixel ASIC and sensor bias, plus detector cooling and temperature monitoring. 
All telescope reference planes share a common clock synchronized to the CERN accelerator timing.
Each measurement starts on a Time-0 reference from the timing telegram, synchronous with the particle spill.

The telescope is orchestrated as a whole by the Constellation framework~\cite{Constellation}, which handles configuration, synchronization, run management, logging, monitoring and notification. 
Each subsystem (reference planes, motion stages and goniometer, system monitoring and database interface) is integrated through a dedicated Constellation satellite software module.
Slow-control telemetry is collected into a PostgreSQL/TimescaleDB database and visualized with Grafana dashboards. 
Track reconstruction, alignment and data analysis use the modular reconstruction and analysis framework Corryvreckan~\cite{Corryvreckan}.
The detector response is modeled with the Monte Carlo framework Allpix\textsuperscript{2}~\cite{Allpix2}, which enables simulation of the expected telescope performance.

\section{Commissioning and Performance}

\begin{figure}[!h]
   \centering
   \includegraphics[width=.88\columnwidth]{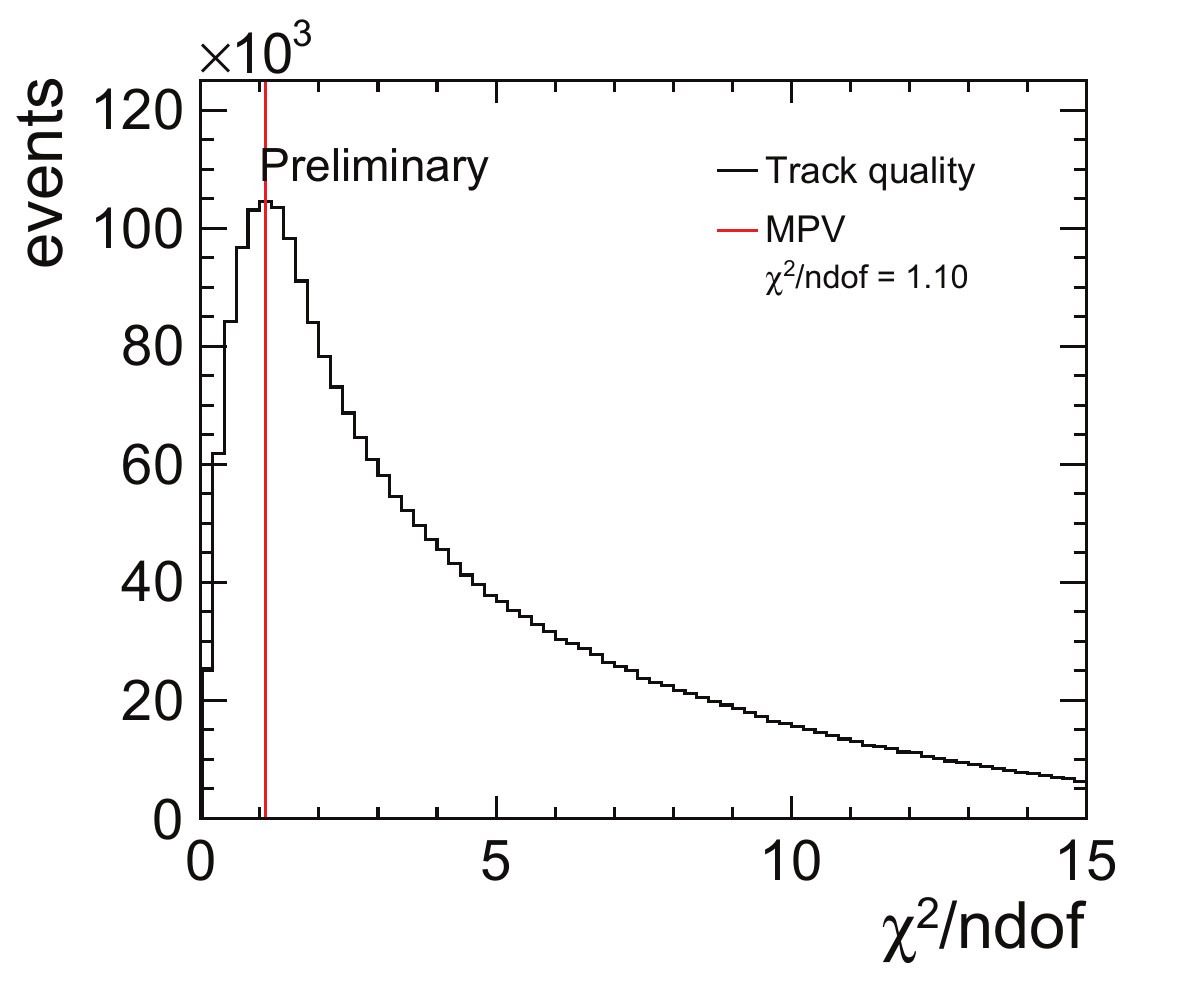}
   \caption{Track-quality/$\chi^2/\mathrm{ndof}$ distribution.}
   \label{fig:chi2}
\end{figure}

The telescope has been commissioned in the H8 beamline and operated in stable, multi-day campaigns.
The reconstruction runs online for monitoring and offline for the final analysis with Corryvreckan.
The quality of the reconstructed tracks is illustrated by the $\chi^2/\mathrm{ndof}$ distribution in Fig.~\ref{fig:chi2}, which peaks at a most-probable value (MPV) of \num{1.10}, confirming a consistent alignment and hit-uncertainty model.

The preliminary tracking performance is summarized in Table~\ref{tab:performance}.
The quoted values are the Gaussian widths of the corresponding distributions and represent upper limits on the underlying resolutions, as they still include the telescope's own contribution and the not-yet-finalized absolute calibration.

\begin{table}[!h]
   \centering
   \caption{Preliminary Tracking Performance (Gaussian Widths)}
   \begin{tabular}{lc}
      \toprule
      \textbf{Observable}                 & \textbf{Gaussian width $\sigma$} \\
      \midrule
      Single-plane residual, $x$          & \qty{1.83}{\um}   \\
      Deflection angle, $x$               & \qty{18.08}{\urad} \\
      Time residual                       & \qty{9.42}{ns}    \\
      Track quality $\chi^2/\mathrm{ndof}$ (MPV) & \num{1.10}  \\
      \bottomrule
   \end{tabular}
   \label{tab:performance}
\end{table}

\begin{figure}[!h]
   \centering
   \includegraphics[width=.88\columnwidth]{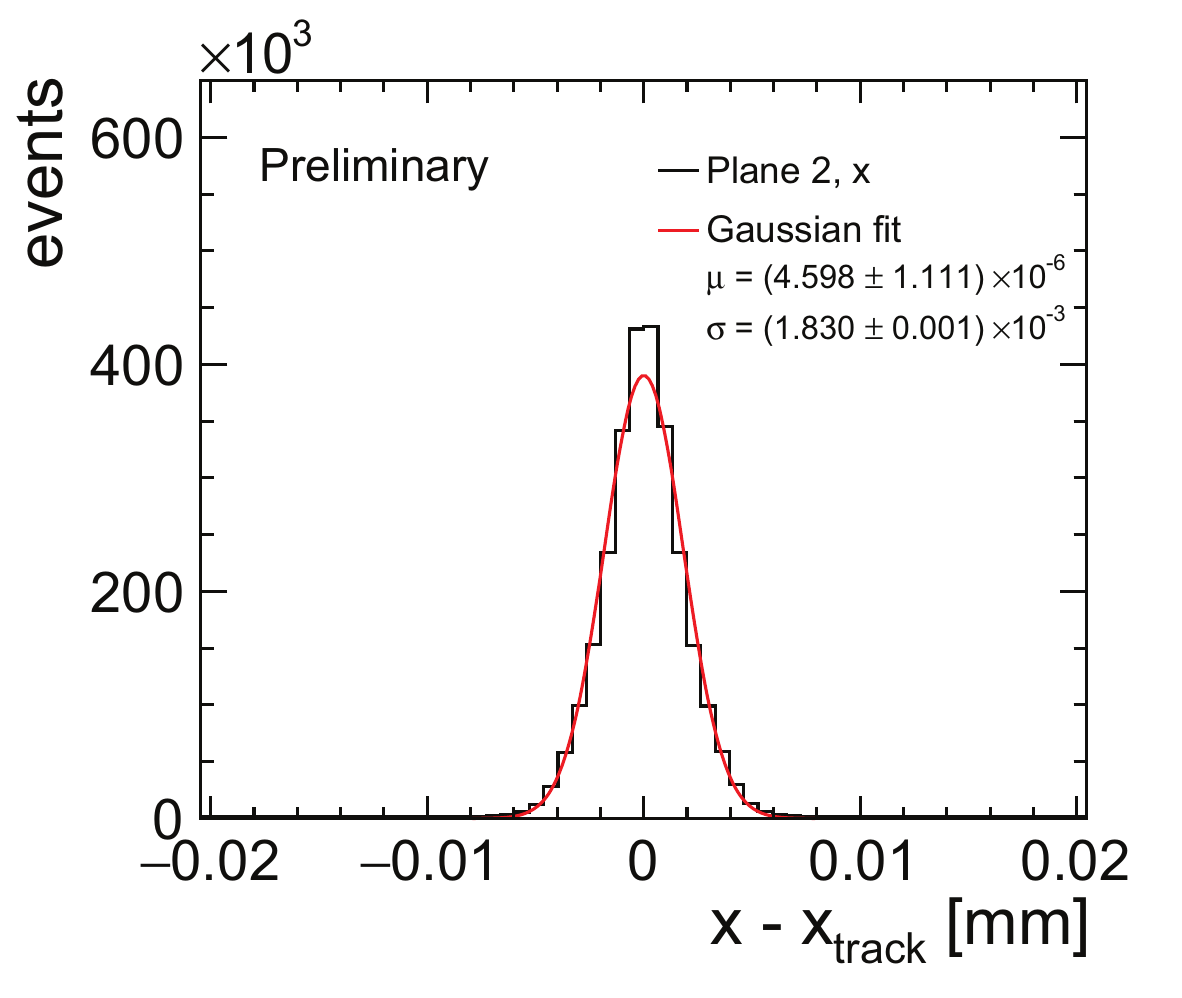}
   \caption{Track residual in $x$ on a reference plane.}
   \label{fig:residual_x}
\end{figure}

Timepix3 reference planes with \qty{300}{\um} planar silicon sensors provide an intrinsic single-hit spatial resolution of $(2.7\pm0.2)\,\unit{\um}$~\cite{VELOtelescope}.
The measured local track residual in $x$ on a reference plane, i.e.\ the pixel-cluster position minus the reconstructed track intercept, has a Gaussian width of \qty{1.83}{\um} (Fig.~\ref{fig:residual_x}).
Since the plane under test enters the track fit, this residual is \emph{biased}: $\sigma_\text{b}^2 = \sigma_\text{int}^2 - \sigma_\text{trk,b}^2$, with $\sigma_\text{trk,b}$ predicted from geometry- and material-budget simulations. 
The extracted intrinsic resolution agrees with the reference value above to within \qty{0.5}{\um}.
The corresponding residual in $y$ has a width of \qty{1.85}{\um}, confirming that the performance is symmetric in the two transverse directions.

\begin{figure}[!h]
   \centering
   \includegraphics[width=.88\columnwidth]{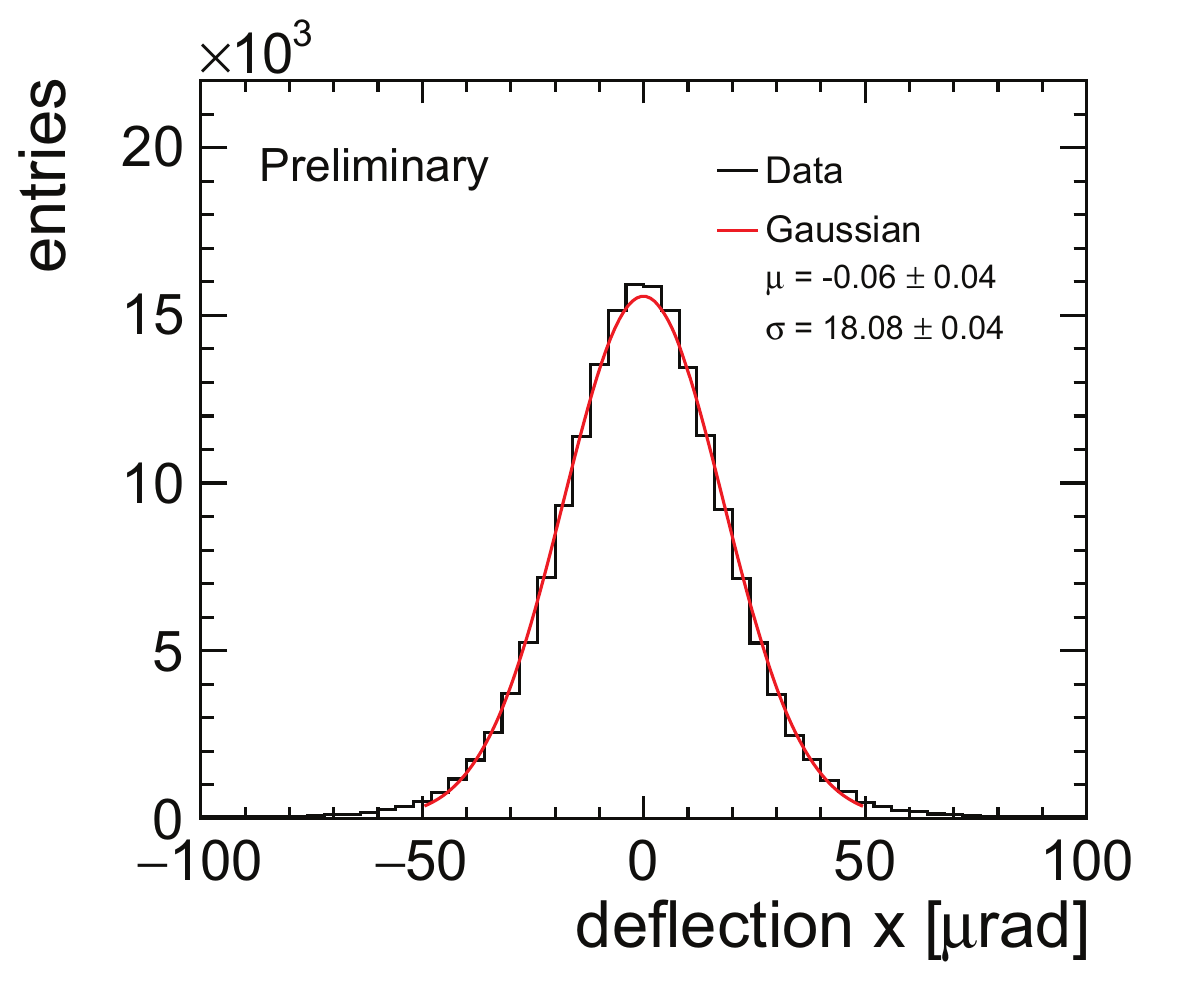}
   \caption{Distribution of the deflection angle in $x$.}
   \label{fig:deflectionx}
\end{figure}

The angular resolution is assessed from the deflection in $x$, i.e. the difference between the upstream and downstream track angles, which has a Gaussian width of \qty{18.08}{\urad} (Fig.~\ref{fig:deflectionx}). 
The deflection in $y$ yields a compatible \qty{18.19}{\urad}, again confirming symmetry in the two directions.
With the sub-micrometre measurement term implied by Eq.~\eqref{eq:angular} and the present arm length, this width is dominated by multiple scattering in the current material budget.

\begin{figure}[!h]
   \centering
   \includegraphics[width=.88\columnwidth]{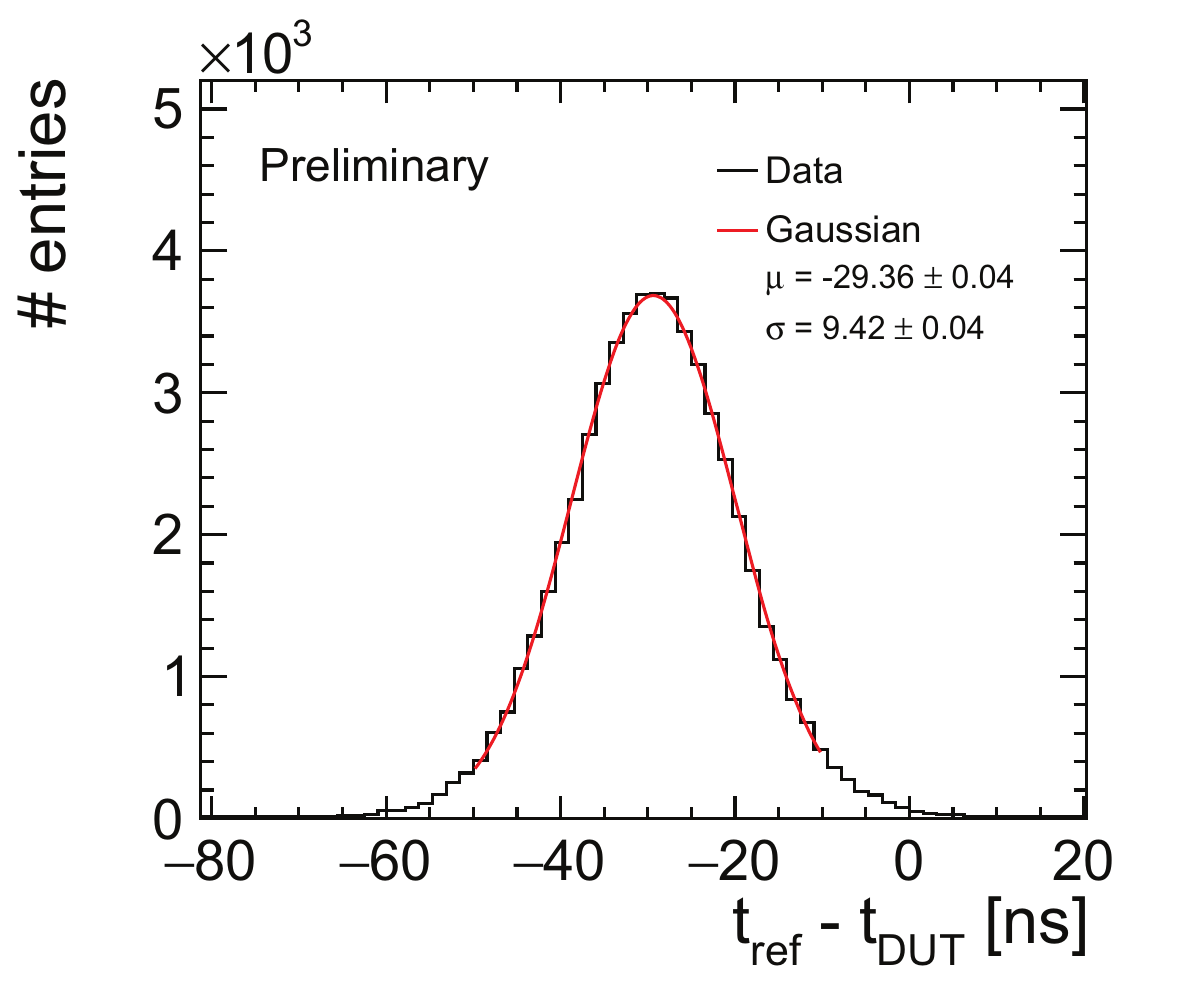}
   \caption{Time residual distribution.}
   \label{fig:residual_time}
\end{figure}

The time residual against the scintillator reference, i.e.\ the scintillator time reference minus the reconstructed track time, has a Gaussian width of \qty{9.42}{ns} (Fig.~\ref{fig:residual_time}).
Its constant offset of about \qty{-29}{ns} is a fixed latency of the trigger path.
The Timepix3 planes themselves reach timing on the order of the on-chip time-to-digital converter bin size~\cite{VELOtelescope}. 
The observed broader residual is dominated by the reference itself.
Its two scintillator--PMT stations sit about \qty{80}{mm} outside the first and last plane and drive long cables to the counting-room front-end, where the coincidence is formed.

\begin{figure}[!h]
   \centering
   \includegraphics[width=\columnwidth]{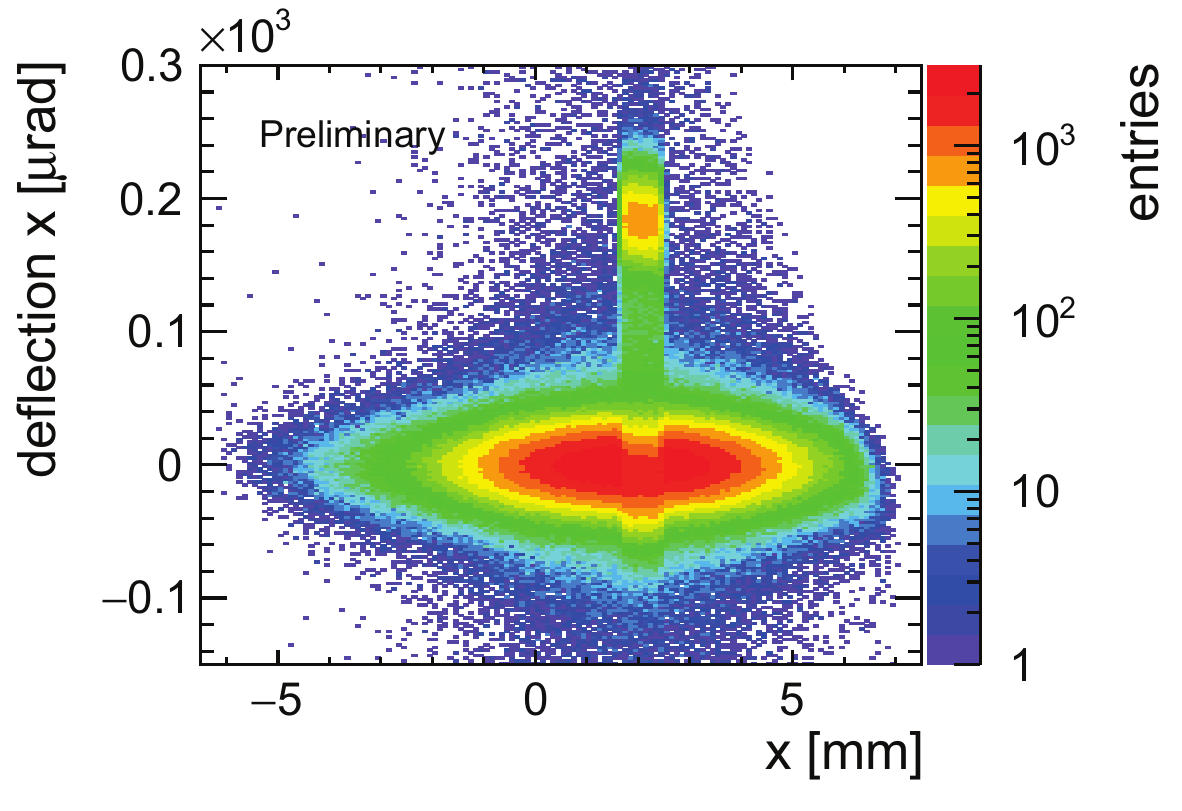}
   \caption{Deflection in $x$ versus track impact position in $x$.}
   \label{fig:deflection}
\end{figure}

With a crystal in beam, the deflection in $x$ is studied as a function of the track impact position on the crystal, shown in Fig.~\ref{fig:deflection}. 
Tracks hitting the \qty{1}{mm}-wide (\qty{1.5}{mm} $< x <$ \qty{2.5}{mm}) crystal face within its angular acceptance are deflected, and the resolved crystal width and deflection spot confirm the setup's suitability for bent-crystal characterization.
These measurements also delineate the improvements required to advance from the present prototype to a fully optimized instrument, motivating the ongoing upgrade.

\section{Ongoing Upgrades}

The principal upgrade is the migration to Timepix4 HPDs (\numproduct{448 x 512} pixels at \qty{55}{\um} pitch, with sub-\qty{200}{ps} timestamp binning)~\cite{Llopart2022, Tpx4Telescope}, targeting the few-\unit{\urad} regime through a substantially reduced material budget. 
This is achieved in three ways: thinning the ASIC from \qtyrange{700}{150}{\um}, using thinner n-on-p sensors (\qty{100}{\um} or \qty{300}{\um}), and replacing the thick multi-layer carrier board behind the chip by a \qty{200}{\um} laminate section from Pyrolytic Graphite Sheet and FR-4.
Together these measures cut the per-plane material budget by up to \qty{90}{\percent}, as detailed in Table~\ref{tab:material-budget}.
The optimized layout for operation after the third CERN Long Shutdown (LS3) places thick outer planes for spatial resolution and thin inner planes for timing resolution, minimizing scattering near the DUT.

\begin{table}[!h]
   \centering
   \caption{Per-Plane Material Content ($X/X_0$) by Reference-Plane Generation, for the \qty{300}{\um} and \qty{100}{\um} Timepix4 Sensor Options}
   \begin{tabular}{lccc}
      \toprule
                          & \textbf{Tpx3}      & \textbf{Tpx4}          & \textbf{Tpx4}          \\
      \textbf{Component}  & \textbf{(current)} & \textbf{\qty{300}{\um}} & \textbf{\qty{100}{\um}} \\
      \midrule
      Sensor              & \qty{0.32}{\percent} & \qty{0.32}{\percent} & \qty{0.11}{\percent} \\
      ASIC                & \qty{0.75}{\percent} & \qty{0.16}{\percent} & \qty{0.16}{\percent} \\
      Carrier / PCB       & \qty{2.69}{\percent} & \qty{0.12}{\percent} & \qty{0.12}{\percent} \\
      \midrule
      \textbf{Total}      & \textbf{\qty{3.76}{\percent}} & \textbf{\qty{0.60}{\percent}} & \textbf{\qty{0.39}{\percent}} \\
      \bottomrule
   \end{tabular}
   \label{tab:material-budget}
\end{table}

The Timepix4 reference planes are designed to be actively cooled by a compact thermoelectric liquid-to-air assembly for temperature-stabilized operation.
A finite-element analysis of the \qty{5}{\watt} HPD predicts a temperature rise of \qtyrange{3}{3.5}{\kelvin} above the coolant, with ample margin from a \qtyrange{50}{100}{\watt} module. 

The readout and control system is upgraded accordingly by extending the capabilities of the BIPXL architecture to the simultaneous readout and control of eight Timepix4 detectors.
This is done together with the corresponding evolution of the Constellation-based control software to accommodate new hardware modules and software features.

\section{Conclusion}

A high-precision beam telescope for the characterization of bent silicon crystals has been designed, developed and commissioned at the CERN SPS North Area. 
The Timepix3-based prototype reconstructs the path of individual beam particles over a configurable \qty{20}{m} baseline with micrometre-level single-plane residuals, an angular resolution of about \qty{18}{\urad} and a per-track time resolution below \qty{10}{ns}.
It has been used for several successful bent-crystal characterization campaigns since spring 2025. 
An in-depth calibration and performance evaluation will follow the ongoing Timepix4 upgrade, which brings a lower material budget, active cooling and a new readout. 
Beyond its present role, the prototype and its upgrade constitute the basis for a possible fleet of beam telescopes serving facilities across CERN. 
Furthermore, the BIPXL telescope addresses an infrastructure need highlighted by the European Committee for Future Accelerators (ECFA) detector R\&D roadmap~\cite{ECFARoadmap}, and, in a reduced-plane-count, adapted-geometry configuration, is a candidate for beam instrumentation at future facilities.

\printbibliography

\end{document}